**Local Ordering in the Pseudogap State of the High-$T_c$ Superconductor Bi$_2$Sr$_2$CaCu$_2$O$_{8+\delta}$**

Michael Vershinin[1,*], Shashank Misra[1,*], S. Ono[2], Y. Abe[2,3], Yoichi Ando[2], and Ali Yazdani[1,†]

[1]Department of Physics and Frederick Seitz Materials Research Laboratory, University of Illinois at Urbana-Champaign, Urbana, Illinois 61801, USA.

[2]Central Research Institute of Electric Power Industry, Komae, Tokyo 201-8511, Japan.

[3]Present address: Correlated Electron Research Center, AIST, Tsukuba 305-8562, Japan.

[*] These authors have contributed equally to this work.

[†] To whom correspondence should be addressed. Currently on sabbatical leave at Geballe Laboratory for Advanced Materials, Stanford University, Stanford CA 94305. email: ayazdani@uiuc.edu

Abstract

We report atomic scale characterization of the pseudogap state in a high-$T_c$ superconductor, Bi$_2$Sr$_2$CaCu$_2$O$_{8+\delta}$. The electronic states at low energies within the pseudogap exhibit spatial modulations having an energy-independent incommensurate periodicity. These patterns, which are oriented along the copper-oxygen bond directions, appear to be a consequence of an electronic ordering phenomenon—the observation of which correlates with the pseudogap in the density of electronic states. Our results provide a stringent test for various ordering scenarios in the cuprates, which have been central in the debate on the nature of the pseudogap and the complex electronic phase diagram of these compounds.

Adding charge carriers to antiferromagetic Mott insulators produces some of the most fascinating and unconventional electronic states of matter. In cuprates, in addition to high-temperature superconductivity, samples with low hole concentration exhibit a conducting state characterized by anomalous transport, magnetic, and optical properties at temperatures above the superconducting transition temperature *(1)*. This regime, the pseudogap state, is characterized by partial gapping of the low energy density of states (DOS), which is detected below a characteristic pseudogap temperature T*, in optical, photoemission, and tunneling spectroscopy measurements *(2)*. It has been argued that the pseudogap state represents a departure from the Fermi liquid picture, and thus requires a fundamentally different electronic description from that used to describe most metals *(3)*. Many of the proposed scenarios involve some form of static or fluctuating ordering phenomena, such as charge, spin *(4-7)* or orbital current order *(8,9)*, as a distinguishing feature of the pseudogap phase. Others have focused on the observation that the pseudogap in underdoped samples evolves continuously into the superconducting gap *(10)*, unusual high frequency conductivity *(11)*, or the Nernst effect in underdoped samples *(12)*, suggesting that the pseudogap state has local pairing correlations required for superconductivity but lacks long-range phase coherence *(13)*. Currently, however, there is no prevailing view on the pseudogap problem.

We report on an atomic scale examination of the electronic states of a cuprate superconductor in its pseudogap state using scanning tunneling microscopy (STM), and show that the electronic excitations with energies below the pseudogap energy exhibit incommensurate spatial modulations that are oriented along the Cu-O bond directions. Among the possible scenarios for the observed ordering, the period of our modulations corresponds most closely to that expected from the pinning of a fluctuating incommensurate spin-density wave *(14,15)*, which is commonly observed in neutron scattering measurements of underdoped cuprates *(16,17)*. However, in addition to spin ordering, we also consider other scenarios as potential explanations for our findings. Regardless of its specific origin, the results establish that the electronic states affected by the pseudogap also exhibit real-space organization.

We also contrast our observation of local ordering in the pseudogap phase with the modulated patterns in the superconducting state. In agreement with previous reports *(18)*, and in strong contrast with our observations in the pseudogap regime, we find the presence of electronic modulations with energy-dependent periodicity in the superconducting state. These modulations in the superconducting state appear for the most part to be described by models of quantum interference of quasiparticle states caused by scattering from random defects *(18-23)*. However, despite the success of the quantum interference picture in describing STM data in the superconducting state, there might be a propensity for ordering *(24,25)* — previous reports on superconducting samples in a magnetic field have shown incommensurate electronic modulations near vortices *(26)*. Currently, it is unclear whether the vortex-induced modulations show any energy dispersion; however, it is conceivable that the local ordering phenomena we have found in the pseudogap state may also nucleate near vortices in the superconducting state *(27)*.

To probe electronic states in both the pseudogap and superconducting states on the atomic scale, we used a home-built ultra-high vacuum (UHV) STM system capable of high-resolution spectroscopic measurements over a wide range of temperatures (from 9K-160K). Our experiments would not have been possible without the high stability of our instrument—keeping registry of the recorded spectroscopic maps with the underlying atomic lattice over a period of several days. To perform STM measurements, the $Bi_2Sr_2CaCu_2O_{8+\delta}$ samples were cleaved *in situ* in UHV at room temperature before being inserted into the cooled microscope stage and stabilized at various temperatures. The measurements reported here were carried out on slightly underdoped $Bi_2Sr_2CaCu_2O_{8+\delta}$ single crystals, which contained 0.6% of Zn impurities ($T_c$ = 80K with pseudogap energy $\Delta_{ps}$ ~ 35-40 meV at 100K, Fig, 1A). Similar results have been observed in Zn-free samples. With samples at 100K, we measured the differential tunneling conductance versus bias voltage and found the presence of a pseudogap in the DOS, with a characteristic pseudogap energy $\Delta_{ps}$ ~ 35-40 meV (Figure 1A). STM topographs of the cleaved surface in the pseudogap state show an atomic corrugation consistent with the BiO surface, together with the structural supermodulation characteristic of the $Bi_2Sr_2CaCu_2O_{8+\delta}$ system (Figure 1B).

The spatial modulation of the pseudogap's electronic state manifests itself when we map the variation of the local DOS at energies below $\Delta_{ps}$. Figures 1C through 1F show unprocessed maps of the differential conductance measured simultaneously at different voltage biases, corresponding to maps of the DOS at specific energies. The DOS at low energies shows periodic patterns oriented along the Cu-O bond directions, which is at a 45° angle relative to the structural supermodulation. To characterize these periodic electronic patterns and determine their connection with the pseudogap in the DOS, we perform a two-dimensional Fourier analysis of the unprocessed conductance maps and track the Fourier amplitude as a function of in-plane wavevector and energy. We show an example of such a Fourier map in Figure 2A, and a schematic labeling of various peaks in this map in Figure 2B. An important distinguishing feature between the Fourier peaks arising from the crystal structure (atomic lattice at A and b-axis supermodulation at S) and those due to the electronic patterns of interest (centered at Q) is their energy dependence (Fig. 2C). The intensity of the structural peaks in the Fourier maps measured at different energies track the overall electronic DOS—a common behavior observed with the STM on various surfaces. However, the Fourier peaks centered at Q show the opposite trend, as they can be detected in the DOS maps only for |eV| approximately less than $\Delta_{ps}$, and more importantly increase in intensity with lowering of energy below $\Delta_{ps}$. The appearance and enhanced intensity of bond-oriented electronic modulations at low energies and their close connection with the pseudogap energy is the main experimental finding of this paper.

Quantitative analysis of the Fourier maps shows that the bond-oriented modulations have an incommensurate periodicity of 4.7± 0.2$a_0$ ($a_0$ is the Cu-Cu distance) for samples containing Zn (Figs. 2A, 2D). While the intensity of the peaks centered at Q varies with energy, their wavelength remains independent of energy within our experimental resolution (± 0.2$a_0$), (Fig. 2D). Measurements on different samples suggest that the periodicity of the modulations corresponding to points Q varies between 4.5$a_0$ and 4.8$a_0$ among the different samples, perhaps because of differences in doping or Zn concentration, the systematic variation of which remains to be investigated. Furthermore, examining the width of the peaks centered at Q shows that the electronic modulations have a correlation length of 4-5 periods (~90Å)—a behavior consistent with the locally disordered or glassy nature of the modulations in the conductance maps (Fig. 1).

The fundamental significance of the modulations of the local DOS in the pseudogap state becomes more evident when we compare them to those observed in the superconducting state. As previously reported *(18,19,24)*, there are several different modulated patterns in the superconducting state; however, unlike those in the pseudogap state, they all display energy-dependent periods. To illustrate this important property, and for the purpose of comparison with the pseudogap patterns, we focus our attention on the bond-oriented modulations. Figure 3 shows our measurements of the wavevectors associated with these modulations on Zn-free samples in the superconducting state at 40K, together with measurements at 4.2K *(18)*. All data in the superconducting state show an agreement in their overall trends and display similar dispersions with energy, which is remarkable given that they result from measurements made at different temperatures and on samples having different concentrations of defects and doping. From these results, we conclude that energy dispersion is a robust feature of bond-oriented modulations in the superconducting state, hence distinguishing them from those in the pseudogap phase.

To interpret our experimental results in the pseudogap state, we consider various possible scenarios for the observation of a periodic modulation in the local DOS recorded by the STM. We believe it unlikely that the wavevectors centered at Q are due to some form of lattice effect, as they behave quantitatively differently from those associated with the crystal structure (Fig. 2C). Another possibility is that the observed patterns are associated with the formation of standing waves caused by the quantum interference of quasiparticles elastically scattered from defects—as previously proposed for the modulations in the superconducting state *(18-23)*. In this scenario, the wavelengths of the electronic patterns are directly related to the wavevectors that connect points on curves of constant electron energy in k-space. This relation has been previously verified by demonstrating that analysis of low temperatures STM and angle-resolved photoemission (ARPES) data in the superconducting state produce approximately the same Fermi surface for $Bi_2Sr_2CaCu_2O_{8+\delta}$ *(18,19)* (see also comparison in Figure 3). To evaluate whether a similar scenario could apply to our results above $T_c$, we have performed model calculations using various electron Green functions that approximate the shape of the constant energy surfaces as measured by ARPES in the normal and pseudogap states *(28,29)*. Qualitatively, none of the calculated interference patterns, some of which are shown in Figure 4, resemble the experimental patterns (Fig. 2). In general, the features in all of the calculated patterns disperse through a range of wavelengths that should be easily resolved in our data (~ 0.4-1.0 $a_0$ over 40mV). The underlying reason for this dispersion is described by the schematic in Figure 4D— which shows how one wavevector potentially relevant for quantum interference changes with energy. The experimental findings in the pseudogap state reported here are thus also quantitatively inconsistent with the quantum interference scenario, simply because the observed modulations' wavelength is independent of energy (to within 0.2 $a_0$) (supporting online text).

One potential method to overcome the shortcomings of the quasiparticle scattering scenario is to introduce a structure factor *(22)* for the scattering centers that acts like a q-space filter—selecting a particular q for all energies up to the pseudogap energy. However, for the model calculations to emulate the experimental data, the form of the structure factor has to explicitly include an *ordered* array of potential wells with an incommensurate periodicity. Therefore, we conclude that our experimental findings in the pseudogap state reported here are inconsistent with a quantum interference scenario, unless we make some form of *ad hoc* assumption about the scattering potential. This discrepancy can not be a consequence of thermal broadening or lifetime effects, which only make the calculated interference patterns more diffuse in momentum space, thus making them even less like the experimental data.

Next, we consider how the observed incommensurate modulation in the pseudogap state might be related to the various proposals for electronic ordering in the cuprates. On general theoretical grounds, the static electronic patterns we observed could be caused by one of the proposed ordered states or could be a consequence of electronic scattering from a fluctuating order that is pinned by disorder in the sample. The first candidate is one-dimensional charge ordering, i.e. stripes, which is predicted to have a spacing of four lattice constants—close to that reported here. However, the electronic patterns that we observe in the pseudogap state appear to be inherently two-dimensional, suggesting either the absence of stripes or requiring fluctuating or highly disordered stripes in two orthogonal directions.

The second possibility is that of local ordering of spins *(4,5,7)*, fluctuations of which are commonly observed in neutron scattering experiments on underdoped samples both above and below $T_c$ *(16,17)*. Recent theoretical work, in connection with both neutron scattering and STM experiments, have proposed that pinning of spin fluctuations by imperfections can give rise to modulations in the local DOS having a period corresponding to half of that of spin correlations *(15)*. This connection between spin and charge modulations appears to be approximately observed if the magnetic modulation wavevectors from neutron scattering are compared with the STM ordering wavevectors reported here for the pseudogap state. Similar arguments have been made in connection with experimental measurements of electronic modulations in the vicinity of vortices *(15,26)*. Despite the attractive features of this scenario, at this time there is no consensus on whether spin fluctuations alone can explain the opening of the pseudogap, which we find to be directly correlated with the observation of the modulations.

Aside from spin and charge ordering scenarios considered in the paper, there are other proposed ordering phenomena, some of which are related to orbital currents in the Cu-O plaquettes *(8,9)*. At this time, the possible connection between these ideas and the local order we observe in the pseudogap state is difficult to assess, as these models either do not anticipate modulations in the local DOS or predict a periodicity different from the one we report here.

In addition to the variety of ordering proposals, a view of the pseudogap state that has received considerable attention is that of a fluctuating superconducting state, with pair correlation persisting well above the resistive transition *(13)*. Within this and related models, it has been shown that a pair-liquid having a fluctuating phase can also demonstrate modulated patterns in its local DOS. These patterns are predicted to show energy-dependent dispersions related to the shape of the Fermi surface *(30)*. This prediction contradicts our finding of dispersion-less modulations in the pseudogap state, hence posing a difficulty for the pre-formed pairs scenario. However, as an alternative to a phase disordered pair-liquid, it might be possible for the pre-formed pairs to localize and form a disordered static lattice. In this picture, assuming that the pseudogap is related to pairing, the pair-lattice modulation could only be detected when probing electronic states at energies smaller than the pseudogap energy. Such a scenario, which combines elements from different proposals on the pseudogap state, has in fact been discussed with regard to vortex cores in a doped Mott insulator *(31)*.

Overall, the connection between spatially organized electronic patterns and the pseudogap reported here supports the theme that the cuprates' complex phase diagram is a result of a competition between various types of ordering phenomena. While we have not clearly identified an order parameter in the pseudogap regime, our findings together with those in the vortex state *(26,27)* suggest that such an order parameter competes or is incompatible with superconductivity.

**Figure Captions**

**Figure 1.** Energy and spatial dependence of the DOS at 100 K. All conductance (DOS) measurements at 100K were taken using a standard ac lockin technique with an initial tunneling current $I_T$, initial sample bias $V_S$, and a bias modulation of 4mV rms. (A) A typical conductance spectrum ($I_T$= 100 pA ,$V_S$= -150 mV) shows a pseudogap in the DOS at the Fermi energy. (B) A typical topograph, taken at a constant $I_T$= 40pA and $V_S$=-150mV, over a 450 Å x 195 Å field of view shows atomic corrugation and the incommensurate supermodulation along the b-axis. Real-space conductance maps recorded simultaneously at (C) 41 mV, (D) 24 mV, (E) 12 mV, and (F) 6 mV demonstrate the appearance and energy evolution of DOS modulation along the Cu-O bond directions. Also evident from these maps is the presence of electronic variations associated with defects (Zn and others) and the dopant inhomogeneity of the material system. The atomic scale contributions of defects to the DOS in the pseudogap state are of great interest; however, we focus our discussion on the periodic modulations.

**Figure 2.** Fourier analysis of DOS modulations. (A) The Fourier transform was taken of an unprocessed conductance map ($I_T$= 40pA, $V_S$=-150mV) acquired over a 380 Å x 380 Å field of view (on a 200-square grid) at 15mV. The unprocessed conductance map has been included in Science's supplementary information. As shown schematically in (B), the FFT has peaks corresponding to atomic sites (colored black and labeled A), primary (at $2\pi/6.8\ a_o$) and secondary peaks corresponding to the b-axis supermodulation (colored cyan and labeled S), and peaks at $\sim 2\pi/4.7\ a_o$ along the $<\pi,0>$ and $<0,\pi>$ directions (colored red and labeled Q). The point $(0, 2\pi/4a_0)$ is also labeled for reference. The energy evolution of these peaks, scaled by their respective magnitudes at 41mV (Q: 70.3 pS, S: 227.7 pS, A: 35.5 pS), is shown in part (C). (D) Two-pixel-averaged FFT profiles were taken along the dashed line in (B) for the DOS measurement at 15mV shown in (A) and measurements acquired simultaneously at 0mV and -15mV. The positions of key peaks are shown by dashed lines and labeled according to their location in (B).

**Figure 3.** Dispersion of bond-oriented modulations in the superconducting state. A set of conductance maps at several energies, with a 380 Å x 380 Å field of view ($I_T$= 75pA, $V_S$=-150mV, on a 128-square grid), was taken simultaneously on a $Bi_2Sr_2CaCu_2O_{8+\delta}$ ($T_c$=88K) sample at 40K. Here we show the energy dependence of the DOS modulation peak in the Fourier maps of the DOS in the superconducting state. The peak position clearly changes as a function of energy (shown in red) and is in good agreement with both the previous measurements at 4.2 K *(18)* and the simple evolution of scattering wavevectors from band structure considerations. *(18)*

**Figure 4.** The quasiparticle scattering picture, in which electron waves scattering off of random defects interfere coherently to form standing waves, qualitatively disagrees with the data in Figure 2. Using a T-matrix formalism *(21-23)* for scattering quasiparticles from a point-like non-magnetic defect of strength 160mV, we first reproduced the dispersion of the peaks in q space seen in low temperature STM data (not shown). As shown in the main part of this figure, we next calculated the scattering density of states for the same impurity in the first Brillouin zone in q space for Fermi Liquid quasiparticles ($\Gamma$= 3 mV) with the $Bi_2Sr_2CaCu_2O_{8+\delta}$ band structure *(28)* at (A) -20meV, (B) +20meV,

and (C) +60meV. The insets show the same calculation at the respective energies using an electronic state which phenomenologically approximates the Fermi arcs seen in ARPES data at T=100K ($\Delta_0$= 40 mV, $\Gamma_0$= 10 mV, $\Gamma_1$= 3 mV, the extent of the arcs in k space was chosen such that a vector q= (0, $2\pi/4.7a_0$) joins the tips of the arcs at the Fermi energy). *(29)* The features seen in these calculated diagrams do not change appreciably for different choices of the broadening parameters, or the strength or nature of the scattering center; here, unusually small values of $\Gamma$ and $\Gamma_1$ have been used to produce sharper q space pictures that more closely resemble the data. (D) Qualitatively, the scattering density of states can be thought of as a statistical weight assigned to the vector (e.g. yellow arrows) that joins two points on curves of constant electron energy in k-space (blue curves), which we show in the main part of the figure for -20 mV (light blue, light yellow), +20 mV (blue, yellow), and +60mV (dark blue, dark yellow) electrons in the Fermi liquid picture. Even assuming scattering occurs only between the tips of the Fermi arcs from the ARPES data *(29)*, the length of the scattering wavevector changes between $2\pi/4.7a_0$ at the Fermi energy (light blue, light yellow) up to $2\pi/5.1a_0$ at 36 meV (dark blue, dark yellow), as shown in the inset.

32. It is our pleasure to acknowledge discussions with J.C. Campuzano, J.C. Davis, E. Fradkin, J.E. Hoffman, A. Kapitulnik, S. Kivelson, M. V. Klein, R. B. Laughlin, A. J. Leggett, D.-H. Lee, K. McElroy, S. Sachdev, Z.-X. Shen, P.W. Phillips, D. Pines, and S.-C. Zhang. Work supported under NSF (DMR-98-75565 & DMR-03-1529632), DOE through Fredrick Seitz Materials Research Laboratory (DEFG-02-91ER4539), ONR (N000140110071), Willet Faculty Scholar Fund, and Sloan Research Fellowship. AY acknowledge support and hospitality of D. Goldhaber-Gordon and K. A. Moler at Stanford.

**Figure 1—Vershinin *et al.***

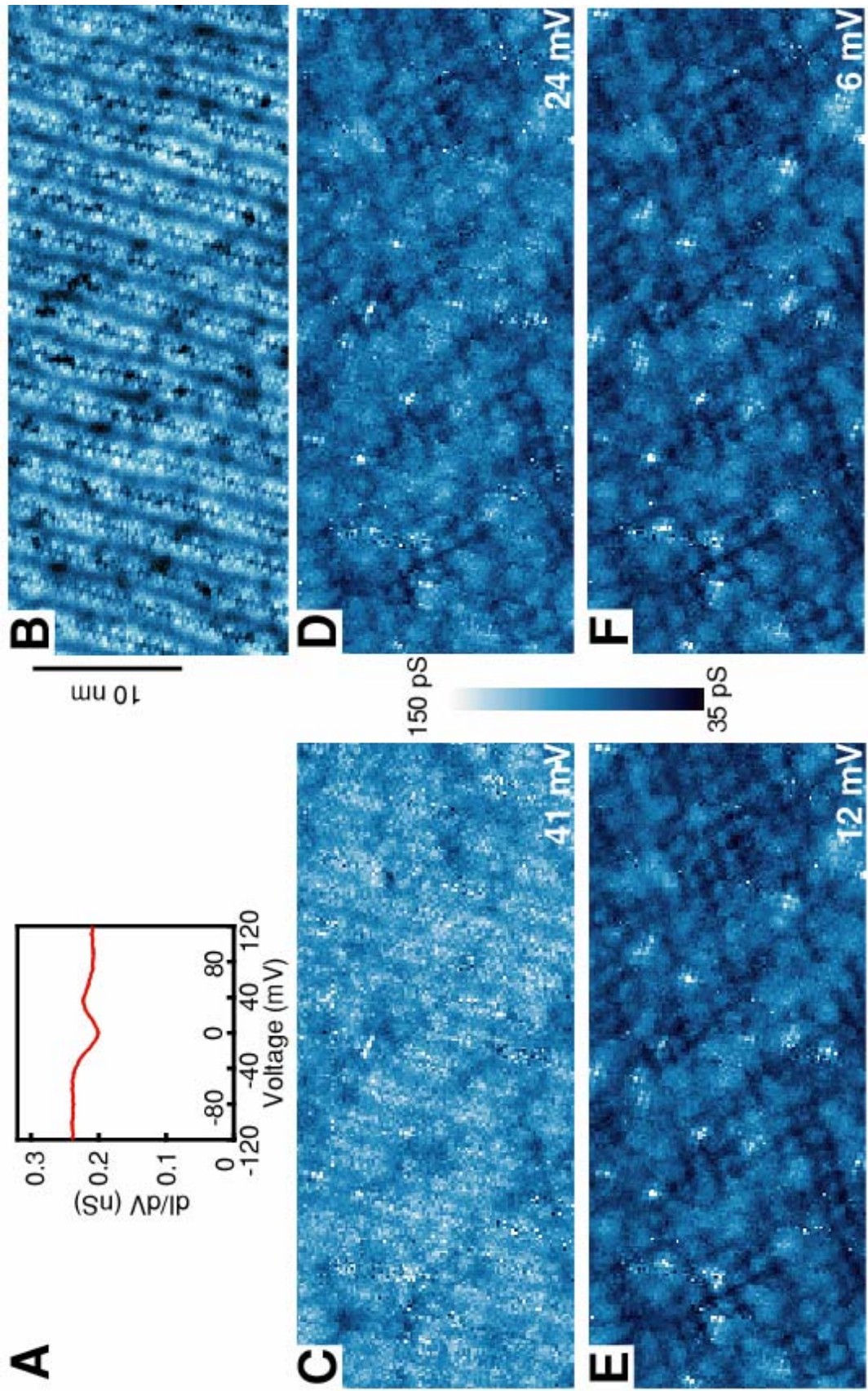

**Figure 2—Vershinin *et al.***

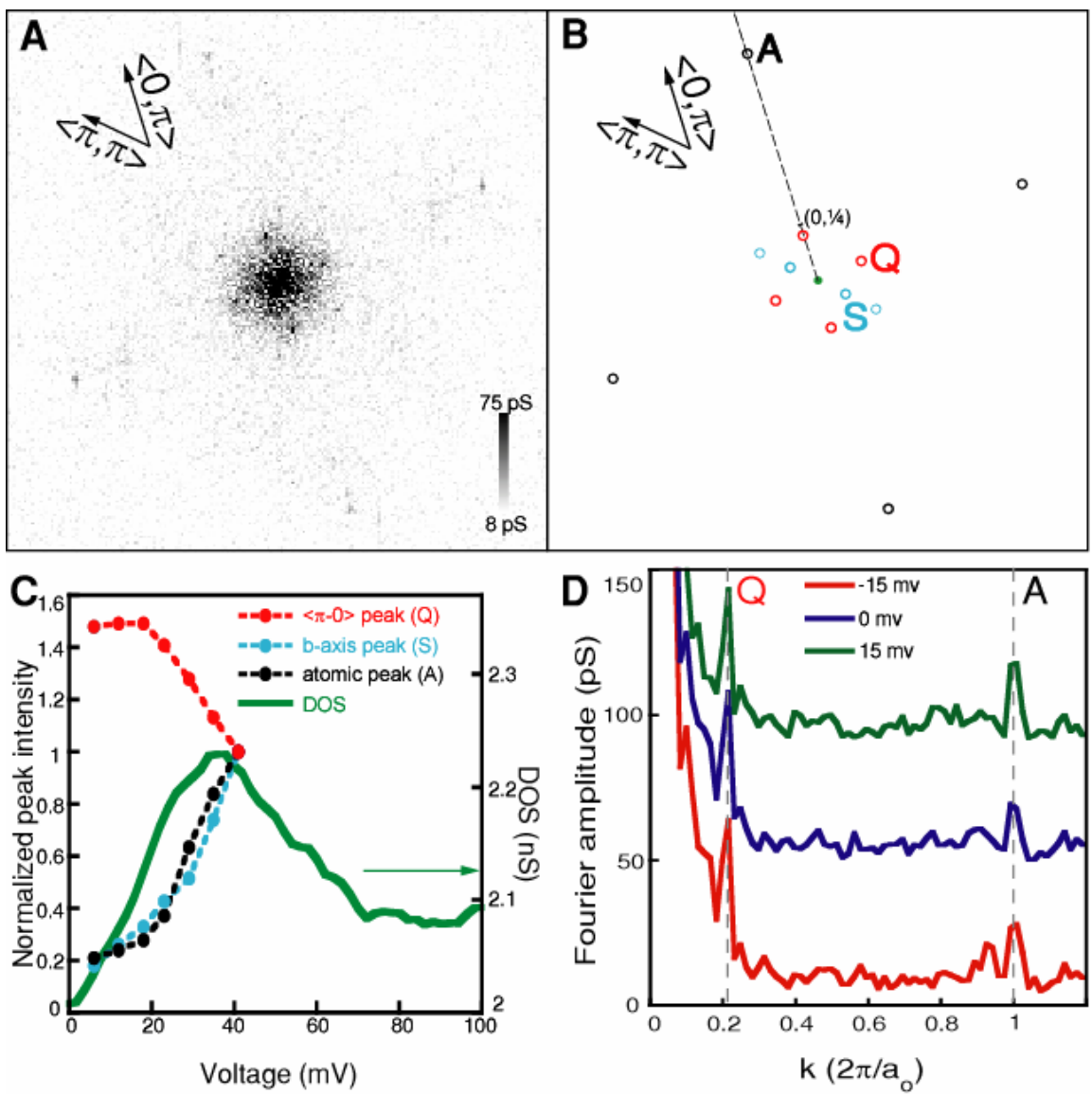

**Figure 3—Vershinin *et al.***

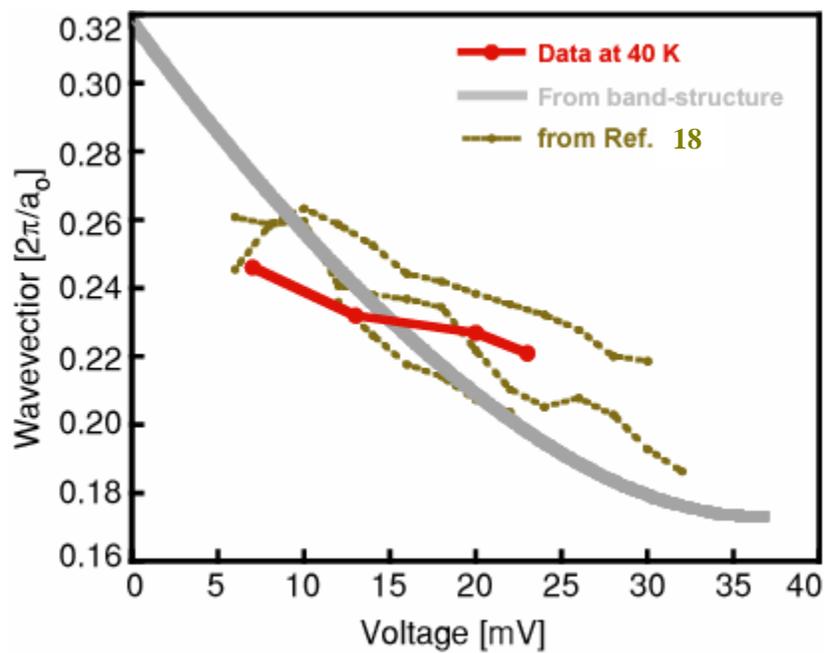

**Figure 4—Vershinin** *et al.*

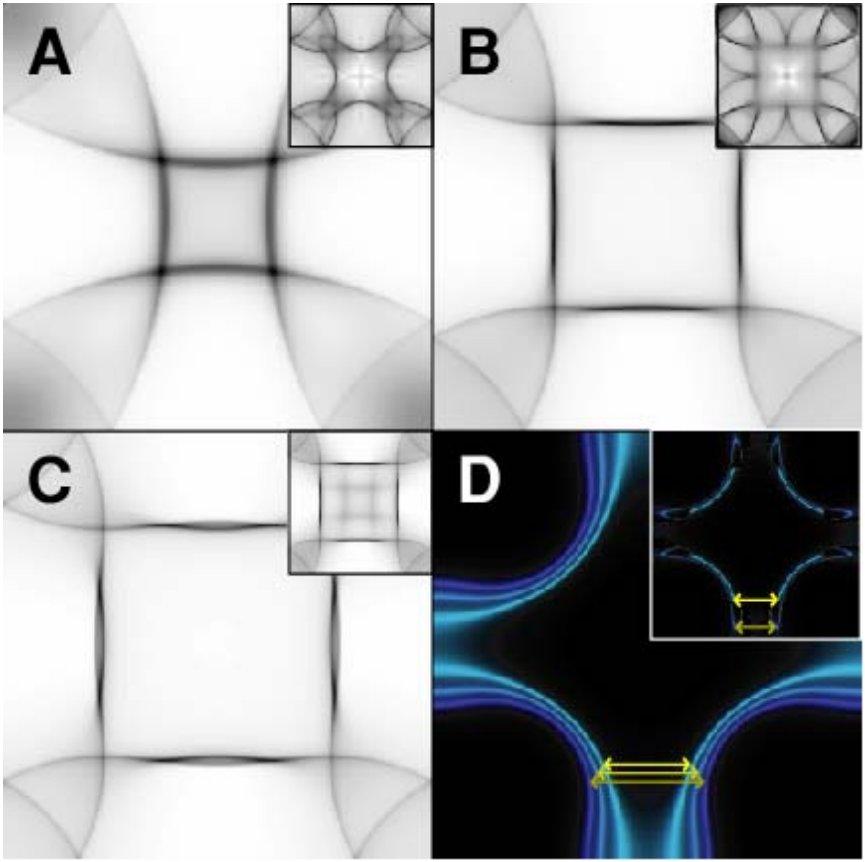